# Ari: The Automated R Instructor

*by Sean Kross, Jeffrey T. Leek, John Muschelli*

**Abstract** We present the `ari` package for automatically generating technology-focused educational videos. The goal of the package is to create reproducible videos, with the ability to change and update video content seamlessly. We present several examples of generating videos including using R Markdown slide decks, PowerPoint slides, or simple images as source material. We also discuss how `ari` can help instructors reach new audiences through programmatically translating materials into other languages.

## Introduction

Videos are a crucial way people learn and they are pervasive in online education platforms (Hsin and Cigas, 2013; Hartsell and Yuen, 2006). Producing educational videos with a lecturer speaking over slides takes time, energy, and usually video editing skills. Maintaining the accuracy and relevance of lecture videos focused on technical subjects like computer programming or data science can often require remaking an entire video, requiring extensive editing and splicing of new segments. We present `ari`, the **A**utomated **R** **I**nstructor as a tool to address these issues by creating reproducible presentations and videos that can be automatically generated from plain text files or similar artifacts. By using **ari**, we provide a tool for users to rapidly create and update video content.

In its simplest form a lecture video is comprised of visual content (e.g. slides and figures) and a spoken explanation of the visual content. Instead of a human lecturer, the **ari** package uses a text-to-speech system to synthesize spoken audio for a lecture. Modern text-to-speech systems that take advantage of recent advancements in artificial intelligence research are available from Google, Microsoft, and Amazon. Many of these synthesizers make use of deep learning methods, such as WaveNet (Van Den Oord et al., 2016) and have interfaces in R (Edmondson, 2019; Muschelli, 2019a; Leeper, 2017). Currently in **ari**, synthesis of the the audio can be rendered using any of these services through the **text2speech** package (Muschelli, 2019b). The default is Amazon Polly, which has text-to-speech voice generation in over twenty one languages, implemented in the **aws.polly** package (Leeper, 2017). In addition to multiple languages, the speech generation services provide voices with several pitches representing different genders within the same language. We present the **ari** package with reproducible use case examples and the video outputs with different voices in multiple languages.

The **ari** package relies on the **tuneR** package for splitting and combining audio files appropriately so that lecture narration is synced with each slide (Ligges et al., 2018). Once the audio is generated, it is synced with images to make a lecture video. Multiple open source tools for video editing and splicing exist; **ari** takes advantage of the FFmpeg (`http://www.ffmpeg.org/`) software, a command-line interface to the `libav` library. These powerful tools have been thoroughly tested with a development history spanning almost 20 years. **ari** has been built with presets for FFmpeg which allow output videos to be compatible with multiple platforms, including the YouTube and Coursera players. These presets include specifying the bitrate, audio and video codecs, and the output video format. The numerous additional video specifications for customization can be applied to command-line arguments FFmpeg through **ari**.

We have developed a workflow with **ari** as the centerpiece for automatically generating educational videos. The narration script for lecture videos is stored in a plain text format, so that it can be synthesized into audio files via text-to-speech services. By storing lecture narration in plain text it can be updated, tracked, and collaboratively or automatically updated with version control software like Git and GitHub. If the figures in the lecture slides are created in a reproducible framework, such as generated using R code, the entire video can be reproducibly created and automatically updated. Thus, **ari** is the Automated R Instructor. We will provide examples of creating videos based on the following sets of source files: a slide deck built with R Markdown, a set of images and a script, or a presentation made with Google Slides or PowerPoint. The overview of the processes demonstrated in this paper are seen in Figure 1. We will also demonstrate the ariExtra package, which contains functions that connect **ari** to applications outside of the R ecosystem (Muschelli, 2020).

## Configuring Ari

Ari relies on several software packages including FFmpeg, one of the most popular libraries for processing audio, video, and image files. Configuring FFmpeg can be challenging, therefore we have provided a Docker image so that Ari users can start producing videos quickly. A guide to getting started with Docker and using our Docker image is included with Ari as a vignette which can be accessed via `vignette("Simple-Ari-Configuration-with-Docker")`. Users who are interested in configuring Ari on their own may find the Dockerfile associated with the guide useful, and it is being actively developed at https://github.com/seankross/ari-on-docker.

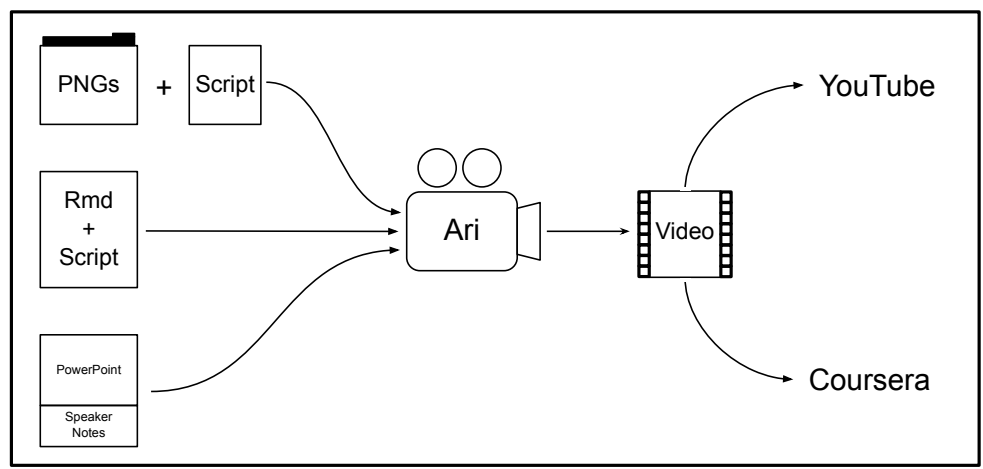

**Figure 1:** Ari is designed to fit into several existing workflows for creating lectures and presentations. Videos can be created with Ari from a series of images and a narrative script, from an R Markdown document, or from a PowerPoint presentation with speaker notes. Ari is pre-configured so that videos are ready to be uploaded to popular platforms like YouTube or Coursera.

### Making videos with **ari**: `ari_stitch`

The main workhorse of **ari** is the `ari_stitch` function. This function requires an ordered set of images and an ordered set of audio objects, either paths to wav files or **tuneR** Wave objects, that correspond to each image. The `ari_stitch` function sequentially "stitches" each image in the video for the duration of its corresponding audio object using FFmpeg. FFmpeg must be installed so that **ari** can combine the audio and images, much like packages such as animation which have a similar requirement (Xie, 2013; Xie et al., 2018b). Moreover, on shinyapps.io, a dependency on the **animation** package will trigger an installation of FFmpeg so **ari** can be used on shinyapps.io. In the example below, 2 images (packaged with **ari**) are overlaid with white noise for demonstration. This example also allows users to check if the output of FFmpeg works with a desired video player.

```
library(tuneR)
library(ari)
result <- ari_stitch(
  ari_example(c("mab1.png", "mab2.png")),
  list(noise(), noise()),
  output = "noise.mp4"
)
isTRUE(result)

[1] TRUE
```

The output indicates whether the video was successfully created, but additional attributes are available, such as the path of the output file:

```
attributes(result)$outfile

[1] "noise.mp4"
```

The video for this output can be seen at https://youtu.be/3kgaYf-EV90.

### Synthesizer authentication

The above example uses `tuneR::noise()` to generate audio and to show that any audio object can be used with **ari**. In most cases however, **ari** is most useful when combined with synthesizing audio using a text-to-speech system. Though one can generate the spoken audio in many ways, such as fitting a custom deep learning model, we will focus on using the aforementioned services (e.g. Amazon Polly) as they have straightforward public web APIs. One obstacle in using such services is that users must go through steps to provide authentication, whereas most of these APIs and the associated R packages do not allow for interactive authentication such as OAuth.

The **text2speech** package provides a unified interface to these 3 text-to-speech services, and we will focus on Amazon Polly and its authentication requirements. Polly is authenticated using the **aws.signature** package (Leeper, 2019). The **aws.signature** documentation provides options and steps to create the relevant credentials; we have also provided an additional tutorial. Essentially, the user must sign up for the service and retrieve public and private API keys and put them into their R profile or other areas accessible to R. Running `text2speech::tts_auth(service = "amazon")` will indicate if authentication was successful (if using a different service, change the `service` argument). NB: The APIs are generally paid services, but many have free tiers or limits, such as Amazon Polly's free tier for the first year (https://aws.amazon.com/polly/pricing/).

**Creating speech from text: `ari_spin`**

After Polly has been authenticated, videos can be created using the `ari_spin` function with an ordered set of images and a corresponding ordered set of text strings. This text is the "script" that is spoken over the images to create the output video. The number of elements in the text needs to be equal to the number of images. Let us take a part of Mercutio's speech from Shakespeare's Romeo and Juliet (Shakespeare, 2003) and overlay it on two images from the Wikipedia page about Mercutio (https://en.wikipedia.org/wiki/Mercutio):

```
speech <- c(
  "I will now perform part of Mercutio's speech from Shakespeare's Romeo and Juliet.",
  "O, then, I see Queen Mab hath been with you.
  She is the fairies' midwife, and she comes
  In shape no bigger than an agate-stone
  On the fore-finger of an alderman,
  Drawn with a team of little atomies
  Athwart men's noses as they lies asleep;"
)
mercutio_file <- "death_of_mercutio.png"
mercutio_file2 <- "mercutio_actor.png"

shakespeare_result <- ari_spin(
  c(mercutio_file, mercutio_file2),
  speech,
  output = "romeo.mp4", voice = "Joanna"
)
isTRUE(shakespeare_result)

[1] TRUE
```

The speech output can be seen at https://youtu.be/SFhvM9gI0kE. We chose the voice "Joanna" which is designated as a female sounding US-English speaker for the script. Each voice is language-dependent; we can see the available voices for English for Amazon Polly at https://docs.aws.amazon.com/polly/latest/dg/SupportedLanguage.html.

Though the voice generation is relatively clear, we chose a Shakespearean example to demonstrate the influence and production value of the variety of dialects available from these text-to-speech services. Compare the video of "Joanna" to the same video featuring "Brian" who "speaks" with a British English dialect:

```
gb_result <- ari_spin(
  c(mercutio_file, mercutio_file2),
  speech,
  output = "romeo_gb.mp4", voice = "Brian"
)
isTRUE(gb_result)

[1] TRUE
```

The resulting video can be seen at https://youtu.be/fSS0JSb4VxM.

The output video format is MP4 by default, but several formats can be specified via specifying the appropriate "muxer" for FFmpeg (see the function `ffmpeg_muxers`). Supported codecs can be founded using the functions `ffmpeg_audio_codecs` and `ffmpeg_video_codecs`. Additional options can be passed to FFmpeg from `ari_stitch` and `ari_spin` to customize the video to the necessary specifications.

We now discuss the number of image and script inputs that **ari** is designed to work with, including text files and a series of PNG images, presentations made with Google Slides or PowerPoint with the script written in the speaker notes section, or an HTML slide presentation created from an R Markdown file, where the script is written in the HTML comments.

### Creating videos from R Markdown documents

Many R users have experience creating slide decks with R Markdown, for example using the [rmark-down](rmark-down) or [xaringan](xaringan) packages (Allaire et al., 2019; Xie et al., 2018a; Xie, 2018). In **ari**, the HTML slides are rendered using [webshot](webshot) (Chang, 2018) and the script is located in HTML comments (i.e. between `<!--` and `-->`). For example, in the file `ari_comments.Rmd` included in **ari**, which is an `ioslides` type of R Markdown slide deck, we have the last slide:

```
x <- readLines(ari_example("ari_comments.Rmd"))
tail(x[x != ""], 4)

[1] "## Conclusion"
[2] "<!--"
[3] "Thank you for watching this video and good luck using Ari!"
[4] "-->"
```

The first words spoken on this example slide are `"Thank you"`. This setup allows for one plain text, version-controllable, integrated document that can reproducibly generate a video. We believe these features allow creators to make agile videos, that can easily be updated with new material or changed when errors or typos are found. Moreover, this framework provides an opportunity to translate videos into multiple languages, a feature that we will discuss in the future directions.

Using `ari_narrate`, users can create videos from R Markdown documents that create slide decks. An R Markdown file can be passed in, and the output will be created using the `render` function from **rmarkdown** (Allaire et al., 2019). If the slides are already rendered, the user can pass these slides and the original document, where the script is extracted. Passing rendered slides allows with the option for a custom rendering script. Here we create the video for `ari_comments.Rmd`, where the slides are rendered inside `ari_narrate`:

```
# Create a video from an R Markdown file with comments and slides
res <- ari_narrate(
  script = ari_example("ari_comments.Rmd"),
  voice = "Kendra",
  capture_method = "iterative"
)
```

The output video is located at [https://youtu.be/rv9fg_qsqc0](https://youtu.be/rv9fg_qsqc0). In our experience with several users we have found that some HTML slides take more or less time to render when using **webshot**; for example they may be tinted with gray because they are in the middle of a slide transition when the image of the slide is captured. Therefore we provide the `delay` argument in `ari_narrate` which is passed to **webshot**. This can resolve these issues by allowing more time for the page to fully render, however this means it may take more time to create each video. We also provide the argument `capture_method` to allow for finely-tuned control of webshot. When `capture_method = "vectorized"`, **webshot** is run on the entire slide deck in a faster process, however we have experienced slide rendering issues with this setting depending on the configuration of an individual's computer. However when `capture_method = "iterative"`, each slide is rendered individually in webshot, which solves many rendering issues, however it causes videos to be rendered more slowly.

In the future, other HTML headless rendering engines (webshot uses PhantomJS) may be used if they achieve better performance, but we have found **webshot** to work well in most of our applications.

With respect to accessibility, **ari** encourages video creators to type out a script by design. This provides an effortless source of subtitles, rather than relying on other services such as YouTube to provide speech-to-text subtitles. When using `ari_spin`, if the `subtitles` argument is `TRUE`, then an SRT file for subtitles will be created with the video.

One issue with synthesis of technical information is that changes to the script are required for Amazon Polly or other services to provide a correct pronunciation. For example, if you want the service to say "RStudio" or "ggplot2", the phrases "R Studio" or "g g plot 2" must be written exactly that way in the script. These phrases will then appear in an SRT subtitle file, which may be confusing to a viewer. Thus, some post-processing of the SRT file may be needed.

### Creating videos from other documents

We created the **ariExtra** (https://github.com/muschellij2/ariExtra) package to augment the core functionality of **ari** by extending it to software applications outside of the R ecosystem. These extensions require many additional dependencies, and considering the significant amount of setup already required for **ari**, we believed that this additional functionality should be in a separate package.

To create a video from a presentation made with Google Slides or PowerPoint, the slides should be converted to a set of images. We recommend using the PNG format for these images. To get the script for the video, we suggest putting the script for each slide in the speaker notes section of that slide. Several of the following features for video generation are in our package **ariExtra**. The speaker notes of slides can be extracted using **rgoogleslides** (Noorazman, 2018) for Google Slides via the API or using **readOffice/officer** (Gohel, 2019; Ewing, 2017) to read from PowerPoint documents. Google Slides can be downloaded as a PDF and converted to PNGs using the **pdftools** package (Ooms, 2019). The **ariExtra** package also has a pptx_notes function for reading PowerPoint notes. Converting PowerPoint files to PDF can be done using LibreOffice and the **docxtractr** package (Rudis and Muir, 2019) which contains the necessary wrapper functions.

To demonstrate this, we use an example PowerPoint is located on Figshare (https://figshare.com/articles/Example_PowerPoint_for_ari/8865230). We can convert the PowerPoint to a PDF, then to a set of PNG images, then we extract the speaker notes.

```
pptx <- "ari.pptx"
download.file(paste0(
  "https://s3-eu-west-1.amazonaws.com/",
  "pfigshare-u-files/16252631/ari.pptx"
),
destfile = pptx
)
pdf <- docxtractr::convert_to_pdf(pptx) # >= 0.6.2
pngs <- pdftools::pdf_convert(pdf, dpi = 300)
notes <- ariExtra::pptx_notes(pptx)
notes
```

```
[1] "Sometimes it's hard for an instructor to take the time to record their lectures.
For example, I'm in a coffee shop and it may be loud."

[2] "Here is an example of a plot with really small axes.  We plot the x versus the y
-variables and a smoother between them."
```

The **ariExtra** package also can combine these processes and take multiple input types (Google Slides, PDFs, PPTX) and harmonize the output. The pptx_to_ari function combines the above steps:

```
doc <- ariExtra::pptx_to_ari(pptx)

Converting page 1 to /var/folders/1s/wrtqcpxn685_zk570bnx9_rr0000gr/T/
/Rtmpo6aD9u/filede6236136195.png... done!
Converting page 2 to /var/folders/1s/wrtqcpxn685_zk570bnx9_rr0000gr/T/
/Rtmpo6aD9u/filede62326b98ef.png... done!

doc[c("images", "script")]

$images
[1] "/private/var/folders/1s/wrtqcpxn685_zk570bnx9_rr0000gr/T/
Rtmpo6aD9u/filede6236058cc5_files/slide_1.png"
[2] "/private/var/folders/1s/wrtqcpxn685_zk570bnx9_rr0000gr/T/
Rtmpo6aD9u/filede6236058cc5_files/slide_2.png"
$script
[1] "Sometimes it's hard for an instructor to take the time to record their lectures.
For example, I'm in a coffee shop and it may be loud."
[2] "Here is an example of a plot with really small axes. We plot the x versus the
y-variables and a smoother between them."
```

This output can then be passed to ari_spin.

We will now demonstrate rendering the video with the "Kimberly" voice while using the divisible_height argument to forcibly scale the height of the images to be divisible by 2. This is required by the x264 (default) codec which we have specified as a preset:

```
pptx_result <- ari_spin(pngs, notes,
  output = "pptx.mp4", voice = "Kimberly",
  divisible_height = TRUE, subtitles = TRUE
)
isTRUE(pptx_result)
```

You can see the output at https://youtu.be/TBb3Am6xsQw. Here we can see the first few lines of the subtitle file:

```
[1] "1"
[2] "00:00:00,000 --> 00:00:02,025"
[3] "Sometimes it's hard for an instructor to"
[4] "2"
[5] "00:00:02,025 --> 00:00:04,005"
[6] "take the time to record their lectures."
```

For Google Slides, the slide deck can be downloaded as a PowerPoint and the previous steps can be used, however it can also be downloaded directly as a PDF. We will use the same presentation, but uploaded to Google Slides. The **ariExtra** package has the function `gs_to_ari` to wrap this functionality (as long as link sharing is turned on), where we can pass the Google identifier:

```
gs_doc <- ariExtra::gs_to_ari("14gd2DiOCVKRNpFfLrryrGG7D3S8pu9aZ")
```

```
Converting page 1 to
/var/folders/zw/l4fv__6n4tnbk3xb31dnbt5m0000gn/T//RtmpphWBAj/filebd69651ed561.png...
done!
Converting page 2 to
/var/folders/zw/l4fv__6n4tnbk3xb31dnbt5m0000gn/T//RtmpphWBAj/filebd694b4b0724.png...
done!
```

Note, as Google provides a PDF version of the slides, this obviates the LibreOffice dependency.

Alternatively, the notes can be extracted using **rgoogleslides** via the Google Slides API, however this requires authentication so we will omit it here. Thus, we should be able to create videos using R Markdown, Google Slides, or PowerPoint presentations in an automatic fashion.

## Summary

The **ari** package combines multiple open-source tools and APIs to create reproducible workflows for creating videos. These videos can be created using R Markdown documents, Google Slides, PowerPoint presentations, or simply a series of images. The audio overlaid on the images can be separate or contained within the storage of the images. These workflows can then be reproduced in the future and easily updated. As the current voice synthesis options are somewhat limited in the tenacity and inflection given, we believe that educational and informational videos are the most applicable area.

## Future directions

The **ari** package is already being used to build data science curricula (Kross and Guo, 2019) and we look forward to collaborating with video creators to augment **ari** according to their changing needs. In the following section we outline possible directions for the future of the project.

Since **ari** is designed for teaching technical content, we plan to provide better support for the pronunciation of technical terms like the names of popular software tools. These names are usually not pronounced correctly by text-to-speech services because they are not words contained in the training data used in the deep learning models that these services are built upon. To address this concern we plan to compile a dictionary of commonly used technical terms and the phonetic phrasing and spelling of these terms that are required to achieve the correct pronunciation from text-to-speech services.

In addition to still images and synthesized voices, we would like to develop new technologies for incorporating other automatically generated videos into lectures generated by **ari**. As computer programming, statistics, and data science instructors we often rely on live coding (Chen and Guo, 2019) to demonstrate software tools to our students. Live coding videos suffer from many of the same problems as other kinds of technical videos as we addressed in the introduction. Therefore we plan to build a system for automating the creation of live coding videos. These videos would also be created using plain text documents like R Markdown. They would integrate synthesized narration with code chunks that would be displayed and executed according to specialized commands that would specify when code should be executed in an IDE like RStudio. These commands could also control which panes and tabs of the IDE are visible or emphasized.

As programmatic video creation software improves, we plan to extend **ari** so it can expand its compatibility with different technologies. For example we believe the heavy reliance on an FFmpeg installation can be mitigated in the future with advances in the **av** package. Though the **av** package has powerful functionality and is currently porting more from libav and therefore FFmpeg, it currently does not have the capabilities required for **ari**. Although third party installation from https://ffmpeg.org/ can be burdensome to a user, package managers such as brew for OSX and choco for Windows provide an easier installation and configuration experience.

Although we rely on Amazon Polly for voice synthesis, other packages provide voice synthesis, such as mscstts for Microsoft and googleLanguageR for Google. We created the **text2speech** package to harmonize these synthesis options for **ari**. Thus, switching from one voice generation service to another simply involves switching the service and voice arguments in **ari**, assuming the service is properly authenticated. This ease of switching allows researchers to compare and test which voices and services are most effective at delivering content.

We see significant potential in how **ari** could expand global learning opportunities. Video narration scripts can be automatically translated into other languages with services like the Google Translation API, where **googleLanguageR** provides an interface. Amazon Polly can speak languages other than English, meaning that one can write a lecture once and generate lecture videos in multiple languages. Therefore this workflow can greatly expand the potential audience for educational videos with relatively little additional effort from lecture creators. We plan to flesh out these workflows so that video creators can manage videos in multiple languages. We hope to add functionality so that communities of learners with language expertise can easily suggest modifications to automatically translated videos, and tooling so suggestions can be incorporated quickly.

*Sean Kross*
*Department of Cognitive Science, University of California, San Diego*
*9500 Gilman Dr.*
*La Jolla, CA 92093*
seankross@ucsd.edu

*Jeffrey T. Leek*
*Department of Biostatistics, Johns Hopkins Bloomberg School of Public Health*
*615 N Wolfe Street*
*Baltimore, MD 21231*
jtleek@jhu.edu

*John Muschelli*
*Department of Biostatistics, Johns Hopkins Bloomberg School of Public Health*
*615 N Wolfe Street*
*Baltimore, MD 21231*
jmusche1@jhu.edu